\documentclass[12pt]{article}
\usepackage{graphicx}
\usepackage{amsmath,amssymb}

\newcommand{\be}{\begin{equation}}
\newcommand{\ee}{\end{equation}}
\def\a{\alpha}
\def\b{\beta}
\def\p{\partial}
\def\v{\varepsilon}
\def\t{\theta}
\def\tr{{\rm tr}}
\def\cN{{\cal N}}
\def\cD{{\cal D}}

\def\bea{\begin{eqnarray}}
\def\eea{\end{eqnarray}}

\def\cN{{\cal N}}
\def\f{\frac}

\def\bD{\bar{D}}

\begin{document}
%\begin{titlepage}

\vspace{1cm}

\begin{center}
{\Large\bf The Background Field Method for $\cN = 2$, $d3$ Super
Chern-Simons-Matter Theories} \vspace{1cm}

{\large\bf I.L. Buchbinder$\,{}^{*}$,
N.G. Pletnev$\,{}^{+}$%,
\\[8pt]
\it\small $^*$Departamento de Fisica, UFJF, Juiz de Fora, MG, Brazil\\

and

\it\small $^*$Department of Theoretical Physics, Tomsk State
Pedagogical University,\\ 634061 Tomsk, Russia\footnote{Permanent
address},
 {\tt email:\ joseph@tspu.edu.ru}
\\[8pt]
$^+$Department of Theoretical Physics, Institute of Mathematics,
630090 Novosibirsk, Russia\\
{\tt email:\ pletnev@math.nsc.ru}\\[8pt]
}
\end{center}
\vspace{0.5cm}

\begin{abstract}
We develop the superfield background field method and study the
effective action in the $\cN=2$, $d3$ supersymmetric
Chern-Simons-matter systems. The one-loop low-energy effective
action for non-Abelian supersymmetric Chern-Simons theory is
computed to order $F^4$ by use of $\cN = 2$ superfield heat kernel
techniques.

\end{abstract}
%\end{titlepage}

%\tableofcontents

\setcounter{equation}{0}
\section{Introduction}

During the last few years, quantum aspects of $d3$ supersymmetric
theories at perturbative level attracted a considerable attention.
This  was inspired by the papers \cite{schwarz04}, \cite{BLG},
\cite{ ABJM}, \cite{n-alg}, \cite{n-quant}, where for an IR
description of stacks of M2-branes  a highly supersymmetric three
dimensional conformal field theories was proposed   in the same
sense as maximally supersymmetric Yang-Mills theory provides an
effective description of stacks of D-branes. Such models are
referred to as the Bagger-Lambert-Gustavsson (BLG) and
Aharony-Bergman-Jafferis-Maldacena (ABJM) theories. ABJM models
defined as three-dimensional $\cN=6$ superconformal $U(N)\times
U(N)$ Chern-Simons-matter theory with level $(k, -k)$. It is
conjectured to describe $N$ M2-branes located at the fixed point of
the $C^4/Z_k$ orbifold in the static gauge.  It is also argued that
the ABJM model is dual to M-theory on $AdS_4 \times S^7/Z_k$ at
large $N$. For $SU(2)\times SU(2)$ gauge group, the $\cN=6$
supersymmetry is enhanced to $\cN=8$ and the ABJM model coincides
with the BLG model. All these new superconformal field theories
involve a non-dynamical gauge field, described by a Chern-Simons
like term in the Lagrangian, which is coupled to matter fields,
parameterizing the degrees of freedom transverse to the worldvolume
of the M2-branes.

The interest to the  extended $d3$ superconformal theories is
connected with two points. First, these theories represent further
evidence of the duality $AdS_4/CFT_3$ \cite{maldac98}. This duality
opens interesting window that allows us to examine the various
properties of condensed matter  (for a review see e.g. \cite{cond}).
Second, these duality also provide novel results about integrability
of the free/planar sector of the $AdS/CFT$ pair of models and
finiteness properties of these superconformal Chern-Simons-matter
theories in the strong coupling regime (for a review see e.g.
\cite{beis}).

It is known for a long time that the quantization  of a membrane
worldvolume theory is very challenging and one of difficulty is the
nonlocality associated with the deformation of membrane without
changing its volume. A quantum supermembrane theory faces a serious
problem of quantum mechanical instability \cite{unstab}. As a
result, a single (quantum mechanical) super - membrane does not make
sense and we get a multi-body problem in its nature, which can be
regarded as the origin of the continuous spectrum. Therefore, from
the field theory side, the action of M2-brane should go away from
the infra-red fixed point to a nonperturbative
Yang-Mills-Chern-Simons system.

One nontrivial test for the BLG and ABJM models as the dual field
theory of M theory is the study of the membrane scattering
amplitude. In the dual gravity description, it can be given by the
effective action of the probe M2-brane due to the large number of
source M2 branes. In the papers \cite{m2d2} the membrane scattering
were examined in the context of the BLG and ABJM models for the
first time. The analysis of  dimensions of the target space from one
loop effective  $v^4$ potential shows that this is consistent with
the calculation from D2-brane super Yang-Mills action. This
phenomenon can be understood as a signal that the membranes
propagating between two membranes always wrap on the one spacial
direction which becomes the compactified direction. Such a result
suggests that the open membrane, described as a perturbation from
the background, always wraps the compactified direction, even when
$k$ is finite and small, and therefore the BLG theory can probe only
remaining ten dimensions. From the analysis of quantum  correction
\cite{m2d2} it was found out the  complete agreement in membrane
scattering dynamics between the results from the ABJM model and
those from the dual supergravity on $AdS_4 \times S^7/Z_k$ in a
specific gauge for worldvolume diffeomorphism. As a result, it was
discovered the following: (i) there is no correction in the $v^2$
term, where $v$ is a probe brane velocity, (ii) $v^4$ term appears
at one-loop, (iii) there should be a non-renormalization theorem, at
least for $\cN = 8$ supersymmetry. It would be very interesting and
instructive to study the above problem using the manifest
supersymmetric and gauge invariant formalism that requires, in its
turn, the development of techniques for calculating the effective
action in frame of superfield background field method.

On the other hand, it is known that multiple M2-branes in eleven
dimensions are reduced to D2-branes in ten dimensions compactifying
one of the transverse direction to the M2-brane. This procedure is
performed by the novel Higgs mechanism proposed in \cite{8},
\cite{8a}. It has been shown in the frame of this mechanism that the
BLG models and the ABJM-like theories are reduced to the super
Yang-Mills theory describing $N$ coincident D2-branes. In this
process the non-dynamical Chern-Simons gauge fields become
dynamical. Since the super Yang-Mills action is the leading order
approximation in string scale $\a'$ expansion of the non-Abelian
Born-Infeld action, it is natural to expect that the BLG and ABJM
theories directly gives rise to a perturbative expansion in terms of
the inverse Yang-Mills coupling constant, or equivalently, in terms
of the inverse vacuum expected value of the Higgs field. As a
result, arising new higher-order action is non-Abelian, plus a
decoupled Abelian degree of freedom.

We want to draw attention to the fact that  similar  ''Higgs''
effect as quantum effect potentially occurs in all  super
Chern-Simons-matter models and even in pure non-Abelian Chern-Simons
theory and its supersymmetric versions. Then the BLG and ABJM
Lagrangians and supersymmetry transformations presented in
\cite{BLG}, \cite{ ABJM} can be thought  as representing the leading
order terms in Planck scale expansion of a (not yet determined)
non-linear M2-brane theory. This circumstance is analogous to the
fact that $\cN=4, D4$ super Yang-Mills theory represents the leading
order terms of the Born-Infeld action, which is believed to describe
the dynamics of coincident D3-branes. Therefore, it would be
interesting to determine the full theory, in which the leading order
terms are the BLG or ABJM Lagrangians. This ambitious program is
similar to non-Abelian supersymmetric extension of the
Born-Infeld-type action in the $\cN=4, D4$ super Yang-Mills quantum
field theory (See as an example of just a few links \cite{ts},
\cite{koe}, \cite{zanon01}, \cite{grasso} from a large list of
references).

The off-shell loop corrections in Chern-Simons-matter theory
attracts much attention since they generate non-trivial quantum
dynamics for classically non-dynamical gauge field (see e.g.
\cite{mck}). The natural way to study these corrections is given by
effective action which can be treated as a method to derive the new,
higher order in strength, gauge invariant and supersymmetric
functionals.

Gauge invariant and manifestly supersymmetric effective action is
constructed on the base of the superfield background field method.
Evaluation of the effective action within the background field
method is often accompanied  by use of proper time or heat kernel
techniques. These techniques allow us to sum up efficiently an
infinite set of Feynman diagrams with increasing number of
insertions of the background fields and to develop a background
field derivative expansion of the effective action in manifestly
gauge covariant way. Precise determination of the effective action
means an exact solution in an appropriate model of quantum field
theory, that of course is impossible in general. Therefore, the
various approximate approaches are used such as the expansion in the
number of loops and an expansion in powers of derivatives. The
coefficients in this series  directly related to the local
geometrical invariants, constructed from the background fields and
their covariant derivatives. In supersymmetric theories such a
procedure allows to find the gauge invariant and supersymmetric
functionals.

As it is known, the most powerful approach to study the quantum
supersymmetric field theories is to make use of an unconstrained
superfield formulation. Unfortunately, such a formulation for the
$\cN = 6,8$ super Chern-Simons-matter theory is not known. The best
what we know up to now is only  $\cN=3$ off-shell formulation on
$\cN = 3, d3$ harmonic superspace \cite{BILPSZ 09}. In such a
formulation three out of six or eight supersymmetries are realized
off-shell while the other three or five are hidden and the
supersymmetry algebra is closed only on shell. The corresponding
superfield actions involve two hypermultiplet superfields in the
bifundamental representations of the gauge groups and two
Chern-Simons gauge superfields corresponding to the left and right
gauge groups. The $\cN = 3$ superconformal invariance allows only a
minimal gauge interaction of the hypermultiplets. Therefore, the
$\cN = 3, d3$ harmonic superspace methods  should be helpful for
these considerations. Alternatively, one can study the effective
action in the $\cN=2$ superspace \cite{BPS}. From the point of view
of $\cN = 2, d3$ supersymmetry, the $\cN = 4, 6, 8$ Chern-Simons and
super Yang-Mills theory describe coupling of the $\cN = 2$ vector
multiplet to the hypermultiplet $\Phi, \bar\Phi$ in the adjoint
representation as well as one or another set of matter
hypermultiplets $ Q,\bar{Q}$ in the bifundamental representation.

The aim of this paper is to construct the background field method
for $\cN = 2$ super Chern-Simons theories, study the effective
action in terms of unconstrained $\cN = 2, d3$ superfields and
calculate of the leading low-energy contributions to the effective
action. Although the various classical and quantum aspects of $\cN =
2, d3$ supersymmetric theories were extensively studied, the
superfield background field method, allowing to develop manifestly
gauge invariant and $\cN=2$ supersymmetric perturbation theory has
not been formulated up to now. Just this problem is solved in the
present paper. As the applications of background field method we
show that in case of pure $\cN=2$ super Chern-Simons,
$\eta$-invariant vanishes, but off-shell contributions to the
effective action have a non-trivial complicated structure. For the
computation of local gauge invariant and manifestly $\cN=2$
supersymmetric contributions we use the procedure previously
proposed in \cite{yung84}, which we generalize to be applied for
superfield theories, and the IR cutoff which is similar to that used
in the work \cite{zanon01}. In our case, the scale of the IR cutoff
will play the role of the Yang-Mills coupling constant and of course
breaks the superconformal invariance. In the case of super
Chern-Simons-matter models the role of the IR cutoff parameter is
played by a  vev of material fields.

The background field method and heat kernel techniques for $\cN = 1,
2$, $d4$ super Yang-Mills theories were well-developed (see
\cite{1001}, \cite{Idea}, \cite{HS}, \cite{echaya} for reviews). In
principle the  $\cN=2, d3$ superfield formalism, is analogical, in
some aspects, to $\cN=1, d4$ superfield formulation. Therefore, we
pay basic attention here only to specific details of quantization,
which are significant namely for $\cN=2, d3$ superfield theories.

This paper is organized as follows. In Section 2 we briefly discuss
the formulation of the super Chern-Simons-matter theories in $\cN=2$
superfield. In Section 3 we formulate the background field method.
In Section 4 we consider a structure of one-loop effective action
and develop the superfield heat kernel technique for its
computation. This technique is then used  to compute the $F^2, F^3,
(\cD F)^2,\ldots, F^4$ terms in the low-energy effective action in
superfield form.

\setcounter{equation}{0}
\section{$\cN = 2, d3$ superfield models}
We start with a brief description of the $\cN = 2, d3$ super
Chern-Simons theory \cite{iv91}, \cite{N2algebra}, \cite{N2algebra2}
. The constrained geometry of $\cN = 2$ supergauge field is
formulated in $R^{3|4}$ superspace with coordinates $z^M=\{x^m,
\t^\a,\bar\t^\a\}$ \footnote{we use superspace conventions of
\cite{BPS}.} in terms of the gauge covariant derivatives
$$
\cD_M\equiv\{\cD_m, \cD_\a, \bar{\cD}_\a\}=D_M+i\Gamma_M^a T^a
$$
where $D_M = (\p_m, D_\a=\p_\t+i\bar\t\not\!\p, \bar{D}_\a
=-\p_{\bar\t}-i\t\not\!\p)$ are the "flat" covariant derivatives and
the gauge connection $\Gamma_M$ takes the values in the Lie algebra
of a compact gauge group. The vector multiplet in three dimensions
is built from one real scalar $\phi$, one complex spinor
$\lambda_\a$, one vector field $A_m$ and one real auxiliary scalar
$D$, all in the adjoint representation of the gauge group. In the
Wess-Zumino gauge the component decomposition for $V$ is given by
\be V=i\t^\a\bar\t_\a \phi +\t^\a\bar\t^\b
A_{\a\b}+i\t^2\bar\t^\a\bar\lambda_\a-i\bar\t^2\t^\a\lambda_\a+\t^2\bar\t^2
D\ee The gauge covariant derivatives obey the superalgebra:
\be\label{alg_cov}\{\cD_\a,\bar\cD_\b\}=-2i\cD_{\a\b}+2i\varepsilon_{\a\b}G,
\quad  \{ \cD_\a,\cD_\b\}=\{\bar\cD_\a,\bar\cD_\b\}=0\ee
$$
[\cD_\rho,\cD_{\a\b}]=\v_{\rho(\a}\bar{W}_{\b)}, \quad
[\bar\cD_\rho,\cD_{\a\b}]=-\v_{\rho(\a}W_{\b)}
$$
$$
[\cD_{\a\b},\cD_{\rho\sigma}]=-i\v_{(\a\rho}F_{\b)
\sigma}-i\v_{(\a\sigma}F_{\b)\rho}.
$$
It follows from (\ref{alg_cov}) that the most general solution to
this algebra in terms of the unconstrained prepotential \cite{1001},
\cite{Idea} is \be \cD_\a=e^{-\Omega}D_\a e^{\Omega}, \quad
\bar\cD_\a=e^{\bar \Omega}\bar{D}_\a e^{-\bar\Omega}, \quad
\Omega=\Omega^a T_a\ee where $\Omega^a$ is an arbitrary complex
superfield. A covariantly chiral and antichiral superfields
$Q_c(z)=e^{\bar\Omega}Q$, $\bar{Q}_c(z)=\bar{Q}e^{\Omega}$ is
defined to be annihilated by these operators,
$$\bar\cD_\a Q_c(z)=0, \quad \cD_\a\bar{Q}_c(z)=0, $$
Hermitian part $\Omega$ defines a potential $e^V=e^\Omega
e^{\bar\Omega}$. In gauge chiral representation one gets
$\cD_A=\{\cD_m, e^{-V}D_\a e^V,\bD_\a\}$ and $Q_c=Q,\,
\bar{Q}_ce^{-V}\bar{Q}.$

The superfield strengths are the linear superfield $G$ and chiral
$W_\a$ and antichiral $\bar{W}_\a$ superfields satisfy the Bianchi
identities \be \bar{W}_\a=\cD_\a G, \quad  W_\a=\bar\cD_\a G,\ee \be
\cD_\a\bar{W}_\b=0, \quad \bar\cD_\a W_\b=0,\quad \cD^\a\cD_\a
G=\bar\cD_\a\bar\cD^\a G=0,\ee \be\cD^\a
W_\a+\bar\cD_\a\bar{W}^\a=0, \quad
\cD_{(\a}W_{\b)}+\bar\cD_{(\a}\bar{W}_{\b)}=-4i\cD_{\a\b}G,\ee \be
F_{\a\b}=\f{1}{8}\{\cD_{(\a}W_{\b)}-\bar\cD_{(\a}\bar{W}_{\b)}\},
\ee \be\cD_\a
W_\b=2F_{\a\b}-i\cD_{\a\b}G+\f12\varepsilon_{\a\b}\cD^\rho
W_\rho,\quad \bar\cD_\a
\bar{W}_\b=-2F_{\a\b}-i\cD_{\a\b}G-\f12\varepsilon_{\a\b}\bar\cD_\rho
\bar{W}^\rho.\ee

In $\cN = 2, d3$ superspace,  the gauge invariant Chern-Simons
action reads in two equivalent forms \cite{iv91} \be\label{class}
S_{CS}=\f{ik}{4\pi}\mbox{tr}\int_0^1 dt\int
d^7z\bar{D}^\a\{e^{-2V}D_\a e^{2V}\}e^{-2V}\p_t e^{2V} \ee
$$=\f{ik}{4\pi}\mbox{tr}\int_0^1 dt\int d^7z{D}^\a\{e^{2V}\bar{D}_\a
e^{-2V}\}e^{2V}\p_t e^{-2V}~.
$$
Here the extra parameter $t$ satisfies the boundary conditions
$V(t=0)=0,\, V(t=1)\equiv V.$  After rescaling the potential as
$V_{new}\equiv 2\sqrt{\f{k}{\pi}} V$ we see that the coupling
constant is $\sqrt{\f{\pi}{k}}$.

The superfield Lagrangian for $N_f$ matter chiral superfields $Q^i$
coupled to non-Abelian $\cN=2$ vector multiplet has the form
\be\label{mat} S_{matter}[V,Q,\bar{Q}]= \mbox{tr}\int d^7z
\sum_{i=1}^{N_f}\bar{Q}^ie^{q_iV} Q^i,\ee where the matter field
$Q^i=\{f^i, \psi^i\}$, with global $\mbox{U}(N_f)$ flavor symmetry,
is in an arbitrary representation $R$ of the gauge group. Such a
$\cN = 2$ theory can be formulated for any gauge group $G$ and
chiral superfields in any representation, with arbitrary
superpotential. The more extended supersymmetric theories can be
formulated using the some sets of $\cN = 2$ superfields. For
example, the $\cN=3$ superconformal theory on the $\cN=2$ language
has $n$ pairs of chiral multiplets $Q^i, \tilde{Q}^i$
($i=1,\ldots,n$), transforming in conjugate representations of the
gauge group, and one chiral superfield $\Phi_a$. The action for the
$\cN=3$ Chern-Simons-matter theory has the form \be
S^{\cN=3}=S^{\cN=2}_{CS} +\int d^7z(\bar{Q}e^V
Q+\tilde{Q}e^{-V}\bar{\tilde{Q}})+[\int
d^5z(-\f{k}{4\pi}\mbox{tr}\Phi^2 +\tilde{Q}^i\Phi_a T^a_{ij}
Q^j)+c.c.]~. \ee Here $\Phi$ is an auxiliary chiral superfield in
the adjoint representation, combined with $V$ to give the $\cN=4$
vector multiplet. The scalar and the auxiliary components of $\Phi$
are combined with the corresponding components of $V$ to form a
triplets under the $SU(2)_R$ symmetry. The R-symmetry also rotates
as doublet the lowest component of the chiral superfield $Q$ and of
its conjugate, antichiral superfield $\bar{Q}$. It should be noted
that the most elegant presentation of a large class of classically
marginal models Chern-Simons-matter with manifestly realized $\cN=3$
off-shell supersymmetry is  provided in the $\cN=3$ harmonic
superspace \cite{BILPSZ 09}.

Maximally supersymmetric theories in $2+1$ dimensions with $SO(8)$
R-symmetry were constructed in \cite{BLG}. These theories have an
interesting property that the closure of the supersymmetry requires
the particular combinations of the gauge group and the matter
content, whereas there is no such restriction for $\cN \leq 3$. The
essential feature of these theories is that the matter fields
$X^IX^I_a T^a, \quad I=1,\ldots,4$ take the values in a metrized
version of the  Lie 3-algebra ${\cal A}_n$: \be [T^a, T^b,
T^c]=f^{abc}_{\ \ \ d}T^d, \quad h^{ab}=\mbox{tr}(T^a,T^b), \ee
where the structure constants $f^{abcd}=f^{abc}_{\ \ \  e}h^{ed}$
are totally antisymmetric in upper indices and are subject to some
basic identity. The gauge field takes values in the Lie algebra
associated with the Lie 3-algebra $A_m=A_{m, ab}t^{ab}$, where the
generators act in the fundamental representation as $(t^{ab})^c_{\
d}=f^{abc}_{\ \ \ d}$. When $h^{ab}$ is positive definite, there is
the only such ${\cal A}_4$ 3-Lie algebra (with $f^{abcd} \varpropto
\varepsilon^{abcd}, h^{ab}=\delta^{ab}$) which satisfies all
reasonable physical requirements. On the associated Lie algebra
there exist two invariant tensors which have the required structure
of a Killing form, namely \be G^{ab,cd}=f^{abcd}, \quad
g^{ab,cd}=f^{abe}_{\ \ \ f}f^{cdf}_{\ \ \ e}.\ee

Extending the BLG model to higher numbers of M2-branes by reducing
the number of supersymmetries led to two generalizations of the
notion of a 3-algebra: the generalized 3-Lie algebras and the
Hermitian 3-algebras \cite{n-alg}. These algebras are used in the
ABJM theories, with $\cN = 6$ supersymmetry and $U(N)\times U(N)$
gauge symmetry and in the ABJ theories \cite{ABJM} with $\cN = 6$
supersymmetry and $U(N) \times U(M)$ gauge symmetry, as well as with
$\cN = 5$ supersymmetry and $Sp(2N) \times O(M)$ gauge symmetry.
Similar theories were constructed in \cite{N5}. A classification of
the possible $\cN = 6$ theories of ABJM-type was presented in
\cite{classif}. In all cases, the underlying 3-bracket is no longer
required to be totally antisymmetric. So that in the case of
theories with $\cN=6$ supersymmetry, the structure constants
$f^{ab\bar{c}\bar{d}}=\mbox{tr}(\bar{T}^{\bar{d}}[T^a,T^b;\bar{T}^{\bar{c}}])$
must satisfy the relations
$f^{ab\bar{c}\bar{d}}=-f^{ba\bar{c}\bar{d}}$ and
$f^{ab\bar{c}\bar{d}}=f^{\star \bar{c}\bar{d} ab}$. The triple
product is also required to satisfy the basic identity.

In order to get a compact form of the Feynman rules, it is
convenient to use the capital Roman letters $A, B,\ldots$ to denote
the indices in associated gauge Lie algebra \cite{BLG},
\cite{n-alg}, \cite{n-quant}. In terms of the gauge
algebra indices, the invariant form is given by\\
$<X,Y>=X^{ab}Y^{cd}f_{abcd}=X^AY^b G_{AB}.$ The structure constants
$F_{ABC}=F_{AB}^{\ \ \ D}G_{DC}$, where  $ F_{AB}^{\ \ \ E}\equiv
C_{ab,cd}^{\ \ \ \ \ \ ef}=2f_{ab [c}^{\ \ \ \ [e}\delta_{d]}^{f]}$,
are totally antisymmetric due to {\it ad}-invariance of $<\cdot
,\cdot>.$ Moreover, it is convenient to use the multi-indices $a i$
combining flavor and 3-algebra indices for $Q^{a i}=Q^I.$ For
example, we have for vertices $<\bar{Q}_i, VQ^i>=Q^I V^A(T_A)_I^{\ \
J}\bar{Q}_J$.

By construction, all these models have at least $\cN=2$
supersymmetry. Higher supersymmetry depends on the underlying
3-algebra and the choices the superpotential. Therefore formally,
the  structure of the effective action in the sector of gauge fields
(without violating the gauge symmetry) should have a universal form.
The difference of effective actions of one model from another is
stipulated by the choice of explicit 3-algebra representations and
relations between various Casimir invariants for such Lie
3-algebras.

\section{ Background field quantization }

We quantize the  $\cN = 2$ super Chern-Simons theory in the
quantum-chiral but background vector representation. As a first step
we split the initial superfields $V, Q, \bar{Q}$ into background $V,
Q, \bar{Q}$ and quantum $v, q, \bar{q}$ parts by the rule
\be\label{split} e^{V}\to e^V e^v~, \quad Q \to Q +q~. \ee Since the
background-quantum splitting for matter superfields is a simple sum,
we will pay the basic attention only on gauge superfield.

The initial infinitesimal gauge transformations can be realized in
two different ways:

(i) background transformations \be\label{btr} e^{ v}\longrightarrow
e^{i\tau}e^{ v}e^{-i\tau}\,,\qquad \cD_M\longrightarrow
e^{i\tau}\cD_M e^{-i\tau}, \ee with a real parameter $\tau=\bar\tau$

(ii) quantum transformations \be\label{qtr}e^v\longrightarrow
e^{i\bar\Lambda}e^ve^{-i\Lambda}\,,\qquad \cD_M
\longrightarrow\cD_M,\ee with background covariantly chiral
parameters, $\cD_\a \bar\Lambda=\bar\cD_\a\Lambda=0.$ For
infinitesimal $\Lambda$ we have (see \cite{1001}) \be \delta
v=L_{\f12 v}\{-i(\bar\Lambda+\Lambda) +\coth L_{\f12
v}i(\bar\Lambda-\Lambda)\}~. \ee Both the superfields $ Q$
($\bar{Q}$) and $ q$ ($\bar{q}$)
 are covariantly chiral (antichiral), $\bar\cD_\a Q=\bar\cD_\a \bar
{q}=0$, where the covariant spinor derivatives act in according with
the  representation of the gauge group. It is worth pointing out
that the form of the background-quantum splitting (\ref{split}) and
the corresponding background and quantum transformations
(\ref{btr}), (\ref{qtr}) are analogous to  the $\cN = 1, d4$
non-Abelian super Yang-Mills model \cite{1001}. Our aim now is to
construct an effective action as a gauge-invariant and $\cN=2$
supersymmetric functional of the background superfield $V$.

The presence of the parameter $t$ in (\ref{class}) is very essential
and the direct integration in (\ref{class}) can be explicitly done
only in the abelian case. However, the first (that noted in
\cite{iv91}) variation of (\ref{class}) and second-order expansion
in powers of quantum field $v$  contain no $t$ integration (modulo a
total spinor derivative) \be\label{expand}S \sim\int_0^1 dt
\p_t(v\bar{D}^\a\Gamma_\a)+\f12\int_0^1 dt \p_t(v\bar{D}^\a\cD_\a
v)+{\cal O}(v^3)\ee It is well known that the linear in $v$ term in
(\ref{expand}) should be dropped when considering the effective
action. The quadratic part $S_2$ of quantum action given in
(\ref{expand}) depends on $V$ via the dependence of $\cD_M$ on
background superfield. Each term in the action (\ref{expand}) is
manifestly invariant with respect to the background gauge
transformations.

We now proceed to the quantization of the theory in a manifest $\cN
= 2$ supersymmetric form. To construct the effective action, we can
use the Faddeev-Popov Ansatz. Within the framework of the background
field method, we should fix only the quantum gauge transformations
(\ref{qtr}) keeping the invariance under the background gauge
transformations. It is convenient to choose the gauge fixing
functions in the form analogous to $\cN =1, d4$ theories: $\bar{f} =
\cD^2v$ , $f = \bar\cD^2v$. These functions are covariantly
(anti)chiral and transform under the quantum gauge transformations
(\ref{qtr}). Therefore the ghost action is the same as in the four
dimensional $\cN = 1$ case \cite{1001}, \cite{Idea}: \be
S_{FP}=\tr\int d^3x d^4\theta\,(b+\bar b) L_{\f12 v}[c+\bar c+\coth
(L_{\f12 v})(c-\bar c)] =\tr\int d^3x d^4\theta\,(\bar b c-b\bar
c)+{\cal O}(v)\,. \ee where $c,\bar{c}, b,\bar{b}$ covariantly
chiral and antichiral superfields. The effective action for pure
Chern-Simons theory is given by the following functional integral
\be e^{i\Gamma_{CS}[V]}=e^{iS_{CS}[V]}\int {\cal D}v{\cal D}b {\cal
D}c\, \delta[f-\bar{\cal D}^2 v]\delta[\bar f-{\cal D}^2 v]
e^{iS_2[V,v]+{\cal O}(v^3)+iS_{FP}}\,. \ee

Unlike in $\cN=1, d4$ case we average this expression with the
following weight (see some details for $\cN=2, d3$ theory in
\cite{N2sgraph new}) \be 1=\int {\cal D}f\cD\bar{f}{\cal
D}\varphi\cD\bar\varphi\, \exp\{\f{i}{2\a}\int d^5z
f^2+\f{i}{2\b}\int d^5\bar{z} \bar{f}^2+i\int d^5z \varphi^2 +i\int
d^5\bar{z} \bar\varphi^2\}, \ee where $\a, \, \b$ are the
gauge-fixing parameters and the anticommuting third ghost superfield
$\varphi$ is background covariantly chiral. As a result, we see that
the $\cN = 2$ super Chern-Simons theory is described within the
background field approach by three ghosts. However, the opposite of
$4d$ case, the Nielsen-Kallosh ghost gives no rise to the effective
action even at one loop level.

Further we will study only one-loop effective action in gauge
superfield sector. In this case it is sufficient to consider, under
the functional integral for $\Gamma_{CS}[V]$, only the quadratic
part of gauge fixed action for quantum fields. Then one gets
\be\label{S_2} S_2+S_{gf}=\f12\mbox{tr}\int d^7z\;
v\f14(\cD^\a\bar\cD_\a+\bar\cD^\a\cD_\a+\f{1}{\a}\cD^\a\cD_\a+
\f{1}{\b}\bar\cD^\a\bar\cD_\a)v \equiv \f12\mbox{tr}\int d^7z\;
v{\cal H}_v v~. \ee  The operator ${\cal H}_{v}$ is defined by Eq.
(\ref{S_2}).

Now we should add the contribution of matter superfields. It is done
by considering the integral over matter quantum fields $q, \bar{q}$
of $e^{iS_{matter}[V,q,\bar{q},v]}$, where
$S_{matter}[V,q,\bar{q},v]$ is obtained from (\ref{mat}) by
background-quantum splitting (\ref{split}). For one-loop
approximation it is sufficient to use
$S_{matter}[V,q,\bar{q},v=0]=S^{(2)}_{hyper}$, As a result, we get
the following representation for the one-loop effective action in
the gauge field sector \be\label{EA}
e^{i\Gamma^{(1)}[{V}]}=e^{iS_{CS}[{V}]}\int {\cal D}v{\cal D}b{\cal
D}c {\cal D}q {\cal D}\bar{q} e^{i\tr\int d^7z\, v {\cal H}_v
v+iS_{FP}+iS^{(2)}_{hyper}} \ee
$$=\mbox{Det}^{-\f12}{({\cal H}_v)} \mbox{Det}^{1}({\cal H}_{FP})
\mbox{Det}^{-\f12}({\cal H}_{hyper})~,$$ where \be {\cal
H}_{FP}=\left(\begin{array}{cc}0 &
\f{1}{16}\cD^2\bar\cD^2\\-\f{1}{16}\bar\cD^2\cD^2 &0
\\\end{array}\right)\delta^{(7)}(z,z')\ee
$$=\left(\begin{array}{cc}0 & -\f{1}{4}\cD^2\\ \f{1}{4}\bar\cD^2 &0  \\\end{array}
\right)\left(\begin{array}{cc}\delta_{-}(z,z') & 0\\0
&\delta_{+}(z,z') \\\end{array}\right).$$ The matter superfield
contributions to the effective action is
\be\Gamma^{(1)}_{hyper}=\f{i}{2}\mbox{Tr}\ln {\cal
H}_{hyper}=\f{i}{2}\mbox{Tr}\ln \left(\begin{array}{cc}0 &
-\f{1}{4}\cD^2\delta_{+}(z,z')\\ -\f{1}{4}\bar\cD^2\delta_{-}(z,z')
&0  \\\end{array} \right) \ee Note that it differs from the
contributions of ghosts only by the sign and choice of the
representation of a gauge group.

\setcounter{equation}{0}
\section{One-loop effective action}
In this section we investigate the off-shell one-loop corrections to
the action for $\cN=2$ super Chern-Simons quantum field theory. It
is well known that the one loop effective action is given in terms
of functional determinants of the differential operators in
quadratic part of action for quantum fields. In the theory under
consideration all these operators are the generalized d'Alembertians
acting on superfields. According to the previous section, there are
three basic d'Alembertians which arise in the covariant supergraphs:
(i) the vector d'Alembertian $\Box_{\rm v}$; (ii) the chiral
d'Alembertian $\Box_{+}$; and (iii) the antichiral d'Alembertian
$\Box_{-}$. The vector d'Alembertian is defined by \be\label{B
v}\Box_{\rm v}={\cal
H}_v^2=\f{1}{16}[-\cD\bar\cD^2\cD-\bar\cD\cD^2\bar\cD+\f{1}{\a\b}\{\cD^2,\bar\cD^2\}-16G^2-\f{8i}{\a}\bar{W}^\a\cD_\a+\f{8i}{\b}
W^\a\bar\cD_\a]\ee
$$=\Box_{cov}+(-1+\f{1}{\a\b})\f{1}{16}\{\cD^2,\bar\cD^2\} +\f{i}{2}(W^\a-\f{1}{\a}\bar{W}^\a)\cD_\a-\f{i}{2}(\bar{W}^\a-\f{1}{\b}W^\a)\bar\cD_\a~,$$
where $\Box_{cov}=\f12\cD^{\a\b}\cD_{\a\b}$ and we have used the
 identities
\be\f18\cD^\a\bar\cD^2\cD_\a=-\Box_{cov}+\f{1}{16}\{\cD^2,\bar\cD^2\}+i\bar{W}^\a\bar\cD_\a+\f{i}{2}(\cD^\a
W_\a)-G^2~,\ee
$$\f18\bar\cD^\a\cD^2\bar\cD_\a=-\Box_{cov}+\f{1}{16}\{\cD^2,\bar\cD^2\}-i{W}^\a\cD_\a-\f{i}{2}(\bar\cD^\a
\bar{W}_\a)-G^2~.$$ It is clear that the most convenient gauge
choice is $\a=\b=1.$

The covariantly chiral d'Alembertian is defined by \be\label{B+}
\Box_{+}= \Box_{cov} +iW^\a  \cD_\a+\f{i}{2}(\cD^\a W_\a) +G^2,\quad
\Box_{+}\Phi=\f{1}{16}\bar\cD^2\cD^2\Phi, \quad \bar\cD_\a\Phi=0.
\ee The antichiral d'Alembertian is defined similarly, \be\label{B-}
\Box_{-}= \Box_{cov} -i\bar{W}^\a  \bar\cD_\a-\f{i}{2}(\bar\cD^\a
\bar{W}_\a) +G^2,\quad
\Box_{-}\bar\Phi=\f{1}{16}\cD^2\bar\cD^2\bar\Phi, \quad
\cD_\a\bar\Phi=0. \ee

Our aim is studying the low-energy effective action $\Gamma[V]$
which is generated by integrating out the quantum fields $v$, ghosts
and matter fields and describes the quantum dynamics of $\cN = 2,
d3$ vector multiplet.

\subsection{Vanishing the $\eta$-invariant}
The operator ${\cal H}_v$ (which is denoted for simplicity as ${\cal
H}$ in this subsection) has the "first order in power $\p$".
Therefore, we must worry about the phase of the functional
determinant. Following the pioneering work \cite{1}, we define the
phase of the path integral by means the superfield eta-invariant of
Atiyach, Patodi and Singer as \be \eta_{\cal H}(s)=\f12 \lim_{s
\rightarrow 0}\sum_{i} \mbox{sign}
\lambda_i|\lambda_i|^{-s}=\mbox{Tr}({\cal H}({\cal
H})^{-\f{s+1}{2}}) \ee Here $\mbox{Tr}$ is a functional trace of
operator acting in superspace. It is evident that the $\eta_{\cal
H}(s)$ is a functional of background field $V$. Then \be\label{eta1}
\f{1}{\sqrt{\mbox{Det}[{\cal H}]}}=\f{1}{\sqrt{\mbox{Det}|[{\cal
H}]|}}e^{i\f{\pi}{4}\eta_{\cal H}(0)}~. \ee In case of
non-supersymmetric Chern-Simons theories the phase in Eq.
(\ref{eta1}) was discussed  in \cite{1}, \cite{5}, \cite{mck}.

Our aim is to compute the $\eta_{\cal H}(0)$ in the theory under
consideration. To do that ones use the identity \be
\label{eta}\eta_{\cal H}(s)=\f{1}{\Gamma(\f{s+1}{2})}\int_0^\infty
dt t^{\f{s-1}{2}}\mbox{Tr}{\cal H}e^{-t{\cal H}^2}~. \ee and then
put $s=0$. For evaluating the integral we, following \cite{Gilkey},
replace the background field $V$ by the field $gV$ with $g$ be a
real parameter. As a result ones get the operator ${\cal H}(g)$,
such that ${\cal H}(1)={\cal H}$ and ${\cal H}(0)$ is background
field independent. Differentiating Eq. (\ref{eta}) one obtains
\be\delta_g\eta_{{\cal
H}(g)}(s)=\f{1}{\Gamma(\f{s+1}{2})}\int_0^\infty dt
t^{\f{s-1}{2}}\mbox{Tr}\{\delta_g{\cal H}(g)e^{-t{\cal H}^2(g)}
-2t\delta_g{\cal H}(g){\cal H}^2(g)e^{-t{\cal H}^2(g)}\}\ee
$$=\f{1}{\Gamma(\f{s+1}{2})}\int_0^\infty dt t^{\f{s-1}{2}}2t^{1/2}\f{d}{dt}\mbox{Tr}\{t^{1/2}\delta_g{\cal H}(g)e^{-t{\cal H}^2(g)}\}$$
Now  we see that $\delta_g\eta_{{\cal H}(g)}$ is regular at $s=0$
and its value is given by a local invariant
\be\label{limeta}\lim_{s\rightarrow 0}\delta_g\eta_{{\cal H}(g)}(s)=
\f{2}{\sqrt{\pi}}\mbox{Tr}(\sqrt{t}\delta_g{\cal H}(g)e^{-t{\cal
H}^2(g)-t\epsilon})|^\infty_0
=-\f{2}{\sqrt{\pi}}\mbox{Tr}(\sqrt{t}\delta_g{\cal H}(g)e^{-t{\cal
H}^2(g)})|_{(t=0)}~.\ee We remind that for given an operator
$\hat{A}$ acting on the space of unconstrained superfields, its
superkernel is determined to be biscalar $A(z,z')$. Then
$\mbox{Tr}\hat{A}$ is given as \be\label{Trac} \mbox{Tr}\hat{A} =
\int
d^3xd^4{\theta}A\delta^3(x-x')\delta^4{(\theta}-{\theta}')|_{x=x',
{\theta}={\theta}'} \ee It is easy to see that the non-zero
contribution in $\delta_g\eta_{{\cal H}(g)}(0)$ can be resulted only
from zero and first terms of power series expansion of
(\ref{limeta}) in $t$ \be \label{Tr} \mbox{Tr} \delta_{g}{\cal
H}e^{-t{\cal H}^2(g)} \sim \f{1}{t^{3/2}}(h_{0}(z)+t h_{1}(z)
+...)~\ee

Then, it is obvious that to obtain a nonzero result, we must put
exactly four spinor derivatives on all Grassmann $\delta$-functions
in Eq. (\ref{Trac}). However, the operator $\delta_{g}{\cal H}(y)$
where $\delta_g\cD_\a=[\cD_\a, e^{-gV}\delta_g e^{gV}],$ $
\delta_g\bar\cD_\a=0$ has a combination of spinor derivatives
$\bar\cD_\a$, $\cD_\a$ of first-degree and the operator ${\cal
H}^{2}(g)$ has also first-order spinor derivatives. Therefore, the
both first terms $h_0(z)$, $h_1(z)$ in (\ref{Tr}) vanish. Thus we
see that $\eta_{{\cal H}(g)}(0)$ does not depend on $g$ and hence it
is background field independent. Therefore this quantity is no more
then  unessential constant and we can omit it in the (\ref{eta1}).
It is known that the background field dependent ${\eta}_{{\cal
H}}(0)$ gives rise to a finite shift of coupling constant in
non-Abelian Chern-Simons action. Our result means that such a shift
is absent in the theory under consideration.

\subsection{Low-energy contribution from vector multiplet}

One-loop effective action, generated by vector multiplet, is given
by the expression \be \Gamma^{(1)}_v[V]=\f{i}{4}\mbox{Tr}\ln {\cal
H}_v^2=\f{i}{4}\mbox{Tr}\int_0^\infty \f{dt}{t}e^{-m^2t}e^{-t{\cal
H}_v^2}\ee where $m$ is infrared regulator. We will calculate the
asymptotic expansion of the heat kernel in the integrand that takes
the form of an expansion in the powers of covariant derivatives.
Structure of such an expansion is defined by superfield DeWitt
coefficient. At the component level, the non-trivial DeWitt
coefficients, $a_n$ for $n \geq 4$, contain in bosonic sector the
field strength terms of the form $F^n$. The first non-trivial
coefficient, $a_4$, is well-known in $d4$ \cite{1001}. In d3 we also
have analogous box diagram with factors $\f{i}{2}{\cal
W}^\a(\cD+\bar\cD)_\a$ at each vertex, and to get non-zero result
one should keep terms with two $D$'s and two $\bar{D}$'s. Besides,
we should treat the gauge strength as matrix in the adjoint
representation $W^\a_{ac}\equiv f_{abc} W^{\a_b}$. Then we get for a
four-points contribution to the effective action: \be
\Gamma^{(1)}_v=-\f{1}{256\pi m^5}\int d^7z g(a_1,a_2,a_3,a_4) ({\cal
W}^\a(a_1) {\cal W}_\a(a_2) {\cal W}^\b (a_3){\cal W}_\b(a_4)\ee
$$ \quad \quad \quad \quad \quad \quad \quad \quad \quad \quad \quad \quad -
\f12{\cal W}^\a(a_1) {\cal W}^\b (a_2){\cal W}_\a(a_3) {\cal
W}_\b(a_4))$$ where in $\cN=2, d3$ case ${\cal W}^\a\equiv
(W-\bar{W})^\a$. Here we have used for colour structures the
notation from work \cite{zanon01}:
$$g(a_1,a_2, \ldots ,a_n)=f_{b_1 a_1 b_2}f_{b_2 a_2
b_3}\ldots f_{b_na_nb_1}~,$$ where $f_{abc}$ are the structure
constants  for a gauge Lie-algebra or $F_{ABC}$ for a gauge
3-algebra. Note that these terms do not have the Abelian analogue.
They simple vanish in the Abelian case.

To obtain the component form and in particular, to get the $\sim
F^4$ terms we should as usual to convert $\int d^4\t
\rightarrow\f{1}{16}D^2\bar{D}^2$ and act with these four spinor
derivatives on the $W_\a=\t^\b f_{\b\a}(x_L)+\ldots$ and
$\bar{W}_\a=\bar\t^\b f_{\b\a}(x_R)+\ldots$, where dots stand for
the terms with derivatives of the fields.

\subsection{The leading contribution of the ghost and matter superfields
} Contribution to one-loop effective action from Faddeev-Popov
ghosts is defined by the expression
$$\mbox{Tr}\ln{\cal
H}_{FP}=\mbox{Tr}\ln\f{1}{16}\bar\cD^2\cD^2+
\mbox{Tr}\ln\f{1}{16}\cD^2\bar\cD^2=\mbox{Tr}_{-}\ln\Box_{-}+
\mbox{Tr}_{+}\ln\Box_{+}$$ Contribution from matter superfields
differs only by sign and choice of representation of gauge group.
That allows to write the ghost contribution to effective action
$\Gamma^{(1)}_{gh}$ in the form of a integral over an auxiliary
proper time $t$ \be i\Gamma^{(1)}_{gh}=\int_0^\infty \f{dt}{t}e^{-t
m^2}(K_{+}(t)+K_{-}(t))~.\ee In this expression, $m^2$ is the
infrared cutoff and $K_{+}(t)$ and $K_{-}(t)$ are the functional
traces of the chiral and antichiral heat kernels respectively, which
are defined by: \be\label{defK}K_{\pm}(t)=\mbox{tr}_{{\cal R}}\int
d^5z_{\pm}(e^{-t\Box_{\pm}}-e^{-t\Box_0})\delta_{\pm}(z,z')|_{z'=z}{\bf
1} =\mbox{tr}_{{\cal R}}\int d^5z_{\pm}K_{\pm}(z,z'|t)|_{z'=z}~.\ee
Here $\mbox{tr}_{{\cal R}}$ denotes the trace over the
representation ${\cal R}$, $dz_{\pm}$ is the integration measure
over (anti)chiral subspace, $\delta_{\pm}(z, z')$ is the
(anti)chiral delta function,
$\delta_{+}(z,z')=-\f14\bar\cD^2\delta^{(7)}(z,z').$ It is
well-known that $K_{+}(t)=K_{-}(t)$. Therefore we discuss only the
computation of the chiral kernel.

One of the procedures in computations of the heat kernel is to make
use of the Fourier integral representation of delta-function as
follows \cite{SK03}: \be \delta^{(7)}(z-z'){\bf 1}=\int
\f{d^3p}{(2\pi)^3}e^{\f{i}{2}\rho^{\a\b}p_{\a\b}}\zeta^2\bar\zeta^2
I(z,z'),\ee where
$$\rho^m=(x-x')^m-i\zeta\gamma^m\bar\t'+i\t'\gamma^m\bar\zeta,
\quad\zeta=\t-\t',\quad \bar\zeta=\bar\t-\bar\t'~.$$ Here $I(z,z')$
is the parallel displacement operator along the geodesic line
connecting the point $z'$ and $z$, defined up to the gauge
transformation $I(z,z')\rightarrow
e^{i\tau(z)}I(z,z')e^{-i\tau(z')}$. For our aims we need only the
following properties of the $I(z,z')$: $I(z,z')I(z',z)=I(z,z)=1$ (as
boundary condition) and the equation $\zeta^A\cD_A
I(z,z')=\zeta^A\cD'_A I(z,z')=0~.$

The heat kernel $K_{+}(z, t)$ has an asymptotic Schwinger-DeWitt
expansion, which is written as \be K_{+}(z, t)=\f{i}{(4\pi
t)^{3/2}}\sum_{n=0}^\infty t^n a_n(z), \quad a_0=a_1=0~. \ee The
$a_n(z)$ are the DeWitt coefficients, which at the component level
contain bosonic field strength terms of the form $F^n$. From
dimensional considerations and the requirement of gauge invariance,
we can expect that the first non-trivial coefficient $a_2$ in the
non-Abelian case is $a_2\sim\mbox{tr}_{{\cal R}}\int d^5z W^2\sim
\mbox{tr}_{{\cal R}}\int d^7z G^2.$ One can show that the $a_n$ with
$n \geq 2$ are obtained in form of $\bar{\cD}^2$ acting on
superfield strengths and their covariant derivatives, and hence all
terms in $K_{+}(z,z'|t)$ can be written as the gauge-invariant
superfunctionals on full superspace. By differentiating the kernel
$K_{+}(z,z'|t)$ with respect to $t$, one observes that: \be\f{d
K_{+}(t)}{dt}=\mbox{tr}_{{\cal R}}\int d^7z \f14\cD^2
e^{-t\Box_{+}}\delta_{+}(z,z')|_{z=z'}~.\ee It is convenient
\cite{grasso} to introduce a new set of coefficients by writing
$$\cD^2 e^{-t\Box_{+}}\delta_{+}(z,z')|_{z=z'}=\f{1}{(4\pi t)^{3/2}}\sum t^n c_n(z)~,$$
as an asymptotic series. Here $a_n(z)=\f{1}{n-\f32}(-\f14\bar\cD^2)
c_{n-1}(z)$. The effective action can then be written as
\be\label{eff}
\Gamma^{(1)}_{gh}=-\f{1}{2\pi^{3/2}}\sum_{n=2}^\infty\f{\Gamma(n-\f32)}{(2n-3)m^{2n-3}}\int
d^7z \mbox{tr}_{\cal R}c_{n-1}.\ee Here ${\cal R}$ means adjoint
representation. Matter contribution has the same form (\ref{eff})
with ${\cal R}$ be a corresponding representation.

Our next goal is to discuss the computations of the superfield
coefficients $c_1$, $c_2$ and $c_3$. We adopt and generalize for
$\cN=2, d3$ case, the procedure developed in \cite{mcart97},
\cite{PB98} and modified for non-Abelian backgrounds in
\cite{grasso}. In some respects, this procedure is similar to that
was used in \cite{yung84} for constructing a gauge invariant
derivative expansion of the effective action in the Yang-Mills
theory.

Using the identities $\int d^2\eta \eta^2=-4$, we present a
$\zeta^2$ as $ \zeta^2=\int d^2\eta e^{\eta^\a\zeta_\a}$. Then
$\f{d}{dt}K_{+}(z, z', t)$ in the point coincidence limit becomes
\be\label{K} K^\a_\a=\int \f{d^3p}{(2\pi)^3}\f14 d^2\eta X^\a X_\a
e^{-t\Delta}\cdot 1~. \ee The operator
$$\Delta=\f12 X^{\a\b}X_{\a\b}+iX^\a W_\a
+G^2~,$$ is defined by $$X_m=\cD_m+ip_m, \quad X^\a=\cD^\a+\eta^\a -
p_{\a\b}\bar\zeta^\b$$ and the auxiliary field $-\f{i}{2}\cD^\a
W_\a$ may be omitted or included in the redefinition $G^2.$  Note
that the last term in $X^\a$ vanishes in the limit of point
coincidence  since the operator $\Box_+$ does not contain
$\bar{\cD}$. Expanding the exponential in powers of proper time
around $e^{p^2t}$ and integrating over $p$, we obtain the desired
expansion, collecting together the coefficients at each degree of
proper time $t$. Due to gauge invariance, these coefficients are
actually expressed in terms of commutators of covariant derivatives.
The pre-exponential factor $X^\a X_\a$ under the integral in
(\ref{K}) is important to write a contribution to effective action
as integral over appropriate superspace. The following integrals are
used:
$$\int\f{d^3p}{(2\pi)^3}e^{tp^2}=\f{i}{(4\pi t)^{3/2}},
\quad \int\f{d^3p}{(2\pi)^3}p_m p_n
e^{tp^2}=\f{-1}{2t}\delta_{mn}\f{i}{(4\pi t)^{3/2}}~,$$
$$\int\f{d^3p}{(2\pi)^3}p_m p_np_l p_k e^{-tp^2}=\f{1}{4t^2}(\delta_{mn}\delta_{kl}+
\delta_{mk}\delta_{nl}+\delta_{ml}\delta_{kn})\f{i}{(4\pi
t)^{3/2}}~. $$

Zeroth-order term does not depend on background fields. In the next
terms of the expansion, we must take into account that $X_\a^3 = 0$
and the integrals over odd powers of $p$ vanishes. Therefore in the
first order of expansion of the (\ref{K}) we have $ -t(\Box+G^2)~,$
with a factor $i/(4\pi t)^{\f32}$ which is common in the expansion.
In the next order of expansion after integration over $p$ we have
exactly $+t\Box$ that cancels gauge non-invariant contribution and
then one gets \be \label{e1}c_1=G^2 \ee as mentioned
above.\footnote{Such a cancelation of gauge non-invariant terms
should take place in any order of heat kernel expansion.} As a
result in the given order of the expansion of the heat kernel, we
obtain the super Yang-Mills action as a leading low-energy
contribution to effective action. The IR cutoff parameter plays a
role of the dimensional coupling constant: \be\Gamma_{gh}^{(1)}
=-\f{1}{4\pi}\mbox{tr}_{ Adj}\int d^5z\f{1}{m}W^\a W_\a.\ee

Discuss some consequence of (\ref{e1}). First, we see that in the
theory under consideration the Chern-Simons action is not induced by
quantum corrections. This conclusion was also pointed out in the end
of subsection 4.1 as a result of vanishing the $\eta$-invariant.
Leading low-energy quantum correction to action is Yang-Mills and
stipulated only by ghosts and matter, vector multiplet does not give
rise. Second, since a contribution of matter to effective action
has, up to a sign, the same form as ghost contribution one can
conclude that for appropriate matter in adjoint representation a
total contribution of ghost and matter to effective action vanishes.
Third, one applies the above consideration to BLG model formulated
in terms of  $\cN =2, d3$ superfields. In this case the ghost and
matter superfields take the values in a real 3-algebra. The induced
Yang-Mills action contains a factor $F_{AC}^{\ \ \ D}F_{BD}^{\ \ \
C}-2(T_A)_I^{\ \ J}(T_B)_J^{\ \ I}=-2G_{AB}$ \cite{n-alg}. Therefore
the leading low-energy correction to action is \be \Gamma_{YM}
=\f{1}{2\pi}G_{AB}\int d^5z\f{1}{m}W^{A\a} W^{B}{}_\a. \ee The
quantities $G_{AB}$ and $W^{A\a}$ are defined in the section 2.

Using the second and third terms in expansion in proper time under
the integral (\ref{K}) one finds the coefficient $c_2$ in the form
\be\label{e2} c_2 = \{\f12
G^4+\f{1}{12}[\cD^m,\cD^n][\cD_m,\cD_n]+\f16[\cD^m,[\cD_m,G^2]]-
\f12[\cD^\a,G^2]iW_\a\ee
$$+\f16[\cD^m,[\cD_m,\cD^\a]]iW_\a+\f13[\cD^m,\cD^\a][\cD_m,iW_\a]\}~.$$
In principle all commutators can be expressed in terms of strengths
and their covariant derivatives. The third term in (\ref{e2})
vanishes since the transfer of the total derivative on the operator
of parallel transport in the limit of coincidence gives $\int
\{\ldots\}\cD_m I(z,z')|_{z=z'}=0.$ To find a component structure of
the coefficient $c_2$ we  use the definition (\ref{alg_cov}) and
Bianchi identities. As a result one gets in bosonic sector the terms
of the form $\sim f^3, (\cD f)^2$ where $f_{mn}$ is bosonic
strength. Note that for Abelian constant background the coefficient
$c_2$ vanishes, since $\mbox{tr}(G\bar{W}^\a W_\a+\bar{W}^\a G
W_\a)=\mbox{tr} G[\bar{W}^\a, W_\a]~.$

The next $c_3$ coefficient in expansion (\ref{eff}) has a
complicated and cumbersome enough structure. To compute it we should
use the expansion of exponential in (\ref{K}) in proper time from
third to six order. The final result for $c_3$ is written as a sum
of two kinds of terms. First, the terms of the form:
 \be\label{1}
-\f{1}{180}\mbox{tr}[-6{\cal O}_1+{\cal O}_2+4{\cal O}_3+3{\cal
O}_4+3{\cal O}_5]~,\ee where \be\label{2}{\cal
O}_1=[\cD_m,\cD_n][\cD_n,\cD_l][\cD_l,\cD_m], \quad{\cal
O}_2=[\cD_m,[\cD_m,\cD_n]][\cD_l,[\cD_l,\cD_n]]~,\ee
$${\cal O}_3=[\cD_m,[\cD_n,\cD_l]][\cD_m,[\cD_n,\cD_l]],\quad {\cal O}_4=
[\cD_m,[\cD_m,[\cD_n,\cD_l]]][\cD_n,\cD_l]~,$$
$${\cal O}_5=[\cD_n,\cD_l][\cD_m,[\cD_m,[\cD_n,\cD_l]]]~.$$
Under the sign of the matrix trace and the integral we have
$\mbox{tr}\int {\cal O}_3=-\mbox{tr}\int {\cal O}_4=-\mbox{tr}\int
{\cal O}_5$ and $\mbox{tr}\int {\cal O}_3=\mbox{tr}\int (2{\cal
O}_2-4{\cal O}_1)$ and then these terms are written as follows:
$$
-\f{1}{180}\mbox{tr}(2F^3-3(\cD F)^2)~. $$ Analogous contribution to
effective action was first constructed in \cite{yung84} for pure
Yang-Mills theory.

Second kind of the terms it is convenient to group according to the
degree $W$. They have  the form: \be\label{3} -\f16 G^6 \ee
\be\label{4} +\f16[G^2, \cD^\a]\{W_\a,\cD^\b\}W_\b-\f16\{[G^2,
\cD^\a],\cD^\b\}W_\a W_\b\ee \be\label{5}
+\f{1}{12}[\cD^\b,\cD_m]\{W_\b,\cD^\a\}[W_\a,\cD_m]+\f{1}{12}[\cD^\b,\cD_m]W_\b[\cD^\a,\cD_m]W_\a\ee
$$+\f{1}{12}\{-2[\cD^\a,\cD_m]\{\cD^\b,[\cD_m,W_\a]\}W_\b-[\cD_m,[\cD^\a,\cD_m]]\{\cD^\b,W_\a\}W_\b+
\{\cD^\b,[\cD_m,[\cD^\a,\cD_m]]\}W_\a W_\b\}$$
$$+\f{i}{6}\{3[\cD^\a, G^2]W_\a G^2-[\cD^\a, G^2][W_\a,G^2]\}$$
\be\label{6} -\f{i}{60}\{-[\cD_m,[\cD_m,\cD^\a]][\cD_n,[\cD_n,
W_\a]]+[\cD^\a,[\cD_n,[\cD_m,\cD_n]]][\cD_m,W_\a]\ee
$$-2[\cD^\a,[\cD_m,\cD_n]][\cD_m,\cD_n]W_\a-3[\cD_m,\cD_n][\cD^\a,[\cD_m,\cD_n]]W_\a\}$$
\be\label{7} -\f{i}{12}\{2[\cD_m,\cD^\a][\cD_m, W_\a]G^2-[\cD^\a,
G^2][\cD_m,[\cD_m, W_\a]]\}\ee \be\label{8}
+\f{1}{12}[\cD_m,G^2][\cD_m,G^2]\ee \be\label{9}
-\f{1}{60}\{[\cD_m,[\cD_m,[\cD_n,[\cD_n,G^2]]]]+[\cD_m,G^2][\cD_n,[\cD_m,\cD_n]]-
[\cD_n,[\cD_m,\cD_n]][\cD_m,G^2]\ee
$$+2G^2[\cD_m,\cD_n][\cD_m,\cD_n]+2[\cD_m,\cD_n][\cD_m,\cD_n]G^2+[\cD_m,\cD_n]G^2[\cD_m,\cD_n]$$
Total coefficient $c_3$ is given by sum of the terms
(\ref{1})-(\ref{9}). In leading bosonic component sector this
coefficient gives us the terms of dimension 8 like $(f_{mn})^4$ and the
products of some power of $f_{mn}$ and some powers of covariant
derivatives $\cD_m f_{nl}$ with total dimension 8.

In principle the coefficient $c_3$ can be transformed to the form
which is expressed completely in terms of superfield strengths and
their supercovariant derivatives, however such an expression will be
extremely tedious and we are not going to write down it there. Note,
that for specific goals and approximations only some certain terms
in $c_3$ can be essential. For example, let us consider the
coefficient $c_3$ for constant Abelian background. Then one can show
that this coefficient is reduced to the following form \be c_3 =
\f{1}{8} \bar{W}^\a \bar{W}_\a W^\b W_{\b}.\ee

As mentioned above, although we considered the heat kernel for the
ghost, the case of the matter chiral superfields can also be treated
along these lines. The results differ from ghost ones only by sign
and choice of representation of gauge group for matter.

\section{Summary and Conclusion}

We have developed the background method for constructing the gauge
invariant effective action in non-Abelian $\cN=2, d=3$
supersymmetric Chern-Simons theory coupled to matter. Using the
background field method we have studied a structure of one-loop
effective action for the theory under consideration. One-loop
effective action was formulated in terms of superfield determinants
of the differential operators on superspace. To evaluate the
determinants we have developed the $\cN =2, d=3$ superfield proper
time technique and formulated a procedure for computing the
low-energy one-loop effective action. It was shown that the leading
quantum correction is defined by ghost and matter superfields. As a
result, the leading contribution in the case of adjoint matter is
$\cN =2, d=3$ Yang-Mills action. A few sub-leading higher derivative
corrections are also calculated.

Background field method opens the possibilities for studying the
effective action in various extended supersymmetric $d=3$ models,
which can be formulated in terms of $\cN=2$ superfields. From our
point of view, a most important application of the methods developed
in the paper is investigation of the effective action in BLG and
ABJM models. These models are the $\cN =8$ and $\cN=6$
supersymmetric Chern-Simons theories coupled to specific matter
supermultiplets. Both models are formulated in terms of $\cN=2$
superfields and hence their quantum aspects can be studied on the
base of the background field method and proper time technique
developed the given paper. However in this case we should consider
the backgrounds containing not only vector multiplet superfield but
also the matter superfields.

\vspace{3mm}
{\bf Acknowledgments}\\[3mm]
The authors are grateful to I.B. Samsonov for collaboration on
earlier stage of work and to A.A. Tseytlin for useful discussions.
The work was partially supported by RFBR grant, project No
09-02-00078 and by a grant for LRSS, project No 3558.2010.2. I.L.B.
is grateful to CAPES for supporting his visit to the Physics
Department of Universidade Federal de Juiz de Fora where the final
part of work was done. Also he acknowledges the support from the
RFBR grants No 11-02-90445. N.G.P.\ acknowledges the support from
RFBR grant, project No 11-02-00242.

\end{document}